\documentclass[aps,prl,twoside,twocolumn,superscriptaddress,floatfix,nofootinbib,showpacs]{revtex4}
\usepackage{amsmath}
\usepackage{bm}
\usepackage{xcolor}

\usepackage{float}
\usepackage{amssymb, amsmath, bm, dcolumn, epsf, graphicx, latexsym, slashed, simplewick}
\usepackage[utf8]{inputenc}
\usepackage[normalem]{ulem}
\usepackage{orcidlink}
\usepackage{soul}
\usepackage{hyperref}
\usepackage{comment}

\newcommand\beq{\begin{equation}}
\newcommand\eeq{\end{equation}}
\newcommand\beqn{\begin{eqnarray}}
\newcommand\eeqn{\end{eqnarray}}

\newcommand{\ba}{\begin{eqnarray}}
\newcommand{\ea}{\end{eqnarray}}
\newcommand{\be}{\begin{equation}}
\newcommand{\ee}{\end{equation}}

\newcommand\lsim{\mathrel{\rlap{\lower4pt\hbox{\hskip1pt$\sim$}}
        \raise1pt\hbox{$<$}}}
\newcommand\gsim{\mathrel{\rlap{\lower4pt\hbox{\hskip1pt$\sim$}}
        \raise1pt\hbox{$>$}}}

\newcommand{\aap}{{Astron.~Astrophys.}}
\newcommand{\apjl}{{Astrophys.~J.~Lett.}}
\newcommand{\apjs}{{Astrophys.~J.~Supp.}}

\usepackage{graphicx}
\usepackage[large]{subfigure}
\usepackage{amssymb, amsmath}
\usepackage[amssymb]{SIunits}
\usepackage{epstopdf}
\usepackage{aas_macros}
\usepackage{natbib}

\begin{document}

%TC:ignore
\title{Did the Universe Reheat After Recombination?}
%TC:endignore

%%%%%%%%%%%%%%%%%%%%%%%%%%%%%%%%%%%%%%%%%%%%%%%%%%%%%%%%%%%%%%%%%%%%%%%%%%%

%TC:ignore
\author{J.~Colin Hill\,\orcidlink{0000-0002-9539-0835}}\email{jch2200@columbia.edu}
\affiliation{Department of Physics, Columbia University, New York, NY 10027, USA}
\affiliation{Center for Computational Astrophysics, Flatiron Institute, New York, NY 10010, USA}
\author{Boris Bolliet}\email{bb667@cam.ac.uk}
\affiliation{DAMTP, Centre for Mathematical Sciences, Wilberforce Road, Cambridge CB3 0WA, UK}
\affiliation{Department of Physics, Columbia University, New York, NY 10027, USA}
%TC:endignore
%%%%%%%%%%%%%%%%%%%%%%%%%%%%%%%%%%%%%%%%%%%%%%%%%%%%%%%%%%%%%%%%%%%%%%%%%%%

%TC:ignore
\begin{abstract}
A key assumption of the standard cosmological model is that the temperature of the cosmic microwave background (CMB) radiation scales with cosmological redshift $z$ as $T_{\rm CMB}(z) \propto (1+z)$ at all times after recombination at $z_\star \simeq 1090$.  However, this assumption has only been precisely tested at $z \lesssim 3$.  Here, we consider cosmological models with post-recombination reheating (PRR), in which the CMB monopole temperature abruptly increases due to energy injection after last scattering.  Such a scenario can potentially resolve tensions between inferences of the current cosmic expansion rate (the Hubble constant, $H_0$).  We consider an explicit model in which a metastable sub-component of dark matter (DM) decays to Standard Model photons, whose spectral energy distribution is assumed to be close to that of the CMB blackbody.  A fit to \emph{Planck} CMB anisotropy, \emph{COBE/FIRAS} CMB monopole, and SH0ES distance-ladder measurements yields $H_0 = 71.2 \pm 1.1$ km/s/Mpc, matter fluctuation amplitude $S_8 = 0.774 \pm 0.018$, and CMB temperature increase $\delta T_{\rm CMB} = 0.109^{+0.033}_{-0.044}$~K, which is sourced by DM decay at $z \gtrsim 10$.  However, matter density constraints from baryon acoustic oscillation and supernovae data highly constrain this scenario, with a joint fit to all datasets yielding $H_0 = 68.69 \pm 0.35$ km/s/Mpc, $S_8 = 0.8035 \pm 0.0081$, and $\delta T_{\rm CMB} < 0.0342$~K (95\% CL upper limit).  These bounds can be weakened if additional dark relativistic species are present in the early universe, yielding higher $H_0$. We conclude that current data disfavor models with significant PRR solely through its impact on background and linear-theory observables, completely independent of CMB spectral distortion constraints.  However, a small amount of such energy injection could play a role in restoring cosmological concordance.
\end{abstract}

% 98.80.-k = cosmology
% 98.80.Es = observational cosmology
% 98.70.Vc = background radiations
% 98.80.Cq = Particle-theory and field-theory models of the early Universe

\maketitle

%TC:endignore

%%%%%%%%%%%%%%%%%%%%%%%%%%%%%%%%%%%%%%%%%%%%%%
\emph{Introduction---}
Recent observations suggest that the current standard model of cosmology, $\Lambda$ cold dark matter ($\Lambda$CDM), may no longer provide an acceptable description of our Universe.  While $\Lambda$CDM provides a good fit to data from cosmic microwave background (CMB) experiments~\cite{Hinshaw2013, Planck2018overview, Aiola2020, SPT-3G:2021eoc}, the predictions of the best-fit model to CMB data are in moderate disagreement with some direct probes of the cosmic expansion rate (the Hubble constant, $H_0$)~\cite{Riess:2019cxk,Riess:2020fzl,Riess:2021jrx,Blakeslee:2021rqi} and the low-redshift matter clustering amplitude (parameterized by $S_8 \equiv \sigma_8 \sqrt{\Omega_\mathrm{m}/0.3}$, where $\sigma_8$ is the root-mean-square linear-theory matter fluctuation amplitude at redshift $z = 0$ and $\Omega_\mathrm{m}$ is the matter density today in units of the critical density)~\cite{DES:2021wwk,2016A&A...594A..24P,Asgari_2021,Krolewski:2021yqy,Dalal:2023olq,Li:2023tui}.  Indeed, the discrepancy in $H_0$ is not unique to CMB data, as large-scale structure (LSS) data analyses yield similar results~\cite{DES:2017txv,eBOSS:2020yzd,Philcox:2020vvt}, lower than direct measurements.  However, the observational situation is not fully clear, as some direct $H_0$ measurements are not in tension with the CMB and LSS constraints~\cite{Freedman:2020dne,Freedman:2021ahq,Boruah:2020fhl,Birrer:2020tax}, and some probes of $S_8$ are not in tension with the CMB~\cite{Planck2018lensing,eBOSS:2020yzd}.  For a comprehensive account of recent measurements, see, e.g., Refs.~\cite{DiValentino:2021izs,Abdalla:2022yfr}.

Nevertheless, the growing number of independent probes yielding similarly anomalous results suggests that the $H_0$ and $S_8$ tensions may indeed reflect a shortcoming of $\Lambda$CDM, and thus a sign of new physics.  In recent years, many scenarios have been explored, with varying degrees of success~\cite{Knox:2019rjx,Schoneberg:2021qvd,DiValentino:2021izs,Abdalla:2022yfr}.  Here we test an underlying assumption of the cosmological model that has evaded attention thus far, motivated by the approach of Ref.~\cite{Ivanov2020}, who showed that CMB data could accommodate a high $H_0$ value by lowering the monopole CMB temperature, $T_\mathrm{CMB,0} \equiv T_{\rm CMB}(z=0)$, from its standard $\Lambda$CDM value (2.7255 K~\cite{Fixsen_2009}) to $T_\mathrm{CMB,0} = 2.56 \pm 0.05 \, \mathrm{K}$ (68\% CL).  This exploits a degeneracy $H_0 \propto T_\mathrm{CMB,0}^{-1.2}$ in $\Lambda$CDM \citep[see][for details]{Ivanov2020}, hinting that the $H_0$ tension can be recast as a tension between CMB anisotropy data combined with local $H_0$ probes and the \emph{COBE/FIRAS} measurement of $T_\mathrm{CMB,0}$ \cite{FIRAS1996,Fixsen_2009}.  Unfortunately, the \emph{FIRAS} error bar on $T_\mathrm{CMB,0}$ is so small ($\approx 1$ mK)~\cite{Fixsen_2009} that no plausible systematic error could bias $T_\mathrm{CMB,0}$ by a sufficient amount to actually resolve the $H_0$ tension in this minimal, theoretically appealing scenario (see also Refs.~\cite{Wen2020,Planck2018parameters} for related discussion).

Motivated by this observation, here we consider a cosmology with lower-than-standard $T_{\rm CMB}(z)$ at $z > z_\star$, where $z_\star \simeq 1090$ is the redshift of recombination. The \emph{FIRAS} measurement of $T_\mathrm{CMB,0}$ is accommodated via a brief epoch of post-recombination ``reheating'' (PRR) during which the usual scaling $T_{\rm CMB}(z) \propto (1+z)$ is violated.  Due to the transient nature of the PRR epoch, this scenario differs from models in which $T_{\rm CMB}(z) \propto (1+z)^\alpha$ with $\alpha \neq 1$, which are tightly constrained by observations at $z \lesssim 3$~(e.g.,~\cite{Saro2014,Luzzi2015,Avgoustidis2016,Li_ACT_2021})   For simplicity, we assume that the injected energy is thermal, but note that PRR scenarios with a wide range of injected photon energies are not ruled out by \emph{FIRAS} or other data~\cite{bcb2021}.  One might expect that a PRR model would decrease the sound horizon at last scattering via earlier recombination due to the lower $T_{\rm CMB}(z \gtrsim z_\star)$, similar in spirit to other modified-recombination proposals~(e.g.,~\cite{Chiang:2018xpn, Sekiguchi:2020teg, Jedamzik:2020krr, Thiele:2021okz}).  However,  the sound horizon is actually \emph{not} decreased relative to $\Lambda$CDM, because of the lower radiation density at early times.  Instead, the PRR scenario yields a higher $H_0$ via an interplay of pre- and post-recombination quantities (see Ref.~\cite{BH23}, hereafter BH23, for details).

To implement the scenario concretely, we consider a model with a sub-component of decaying (or de-exciting) cold dark matter (DCDM) that produces photons at some redshift $z_X$, with $z_X < z_\star$ (``DCDM$\gamma$'' hereafter).  The decay is assumed to take place prior to the era in which direct constraints on $T_{\rm CMB}(z)$ exist (i.e., $z_X \gtrsim 3$).\footnote{Note that in, e.g., decaying axion models the decay rate can be significantly enhanced due to stimulated emission associated with the temperature of an ambient photon bath, which could play a role here~\cite{Caputo:2018vmy}.}  The decay products are assumed to be thermal, such that no significant CMB spectral distortion is generated; we discuss this assumption further below and demonstrate in BH23 that viable parameter space exists.  Regardless, the constraints that we obtain effectively follow directly from the agreement of the \emph{FIRAS} measurement of $T_{\rm CMB,0}$~\cite{Fixsen_2009} and the indirect inference of $T_{\rm CMB,0}$ from (non-\emph{FIRAS}) cosmological datasets~\cite{Ivanov2020}, independent of any spectral distortion signatures.  The relevant mass range of the DCDM is sufficiently low ($\lesssim$ eV) in such models that the metastable component could yield an additional contribution to $N_{\rm eff}$ (the effective number of relativistic species) at $z_\star$; we thus consider a model extension in which $N_{\rm eff}$ is also a free parameter.

%%%%%%%%%%%%%%%%%%%%%%%%%%%%%%%%%%%%%%%%%%%%%%
\emph{Theory---}
The evolution of the background (average) DCDM energy density, $\rho_{\rm DCDM}$, and the photon energy density, $\rho_\gamma$, in our model are, respectively:
\ba
{\rho}^\prime_{\rm DCDM} & = & -3aH\rho_{\rm DCDM}-a\Gamma \rho_{\rm DCDM} \label{eq.background.DCDM} \\
{\rho}^\prime_{\gamma} & = &-4aH\rho_{\gamma} + a\Gamma \rho_{\rm DCDM} \,, \label{eq.background.gamma}
\ea
where $a = (1+z)^{-1}$ is the cosmic scale factor, primes denote derivatives with respect to conformal time $\tau$ (defined by ${\rm d}\tau = {\rm d}t/a$ where $t$ is the cosmic time), $H(a) = ({\rm d}a/{\rm d}t)/a$ is the Hubble parameter defined with respect to cosmic time (thus $aH = a^\prime/a$), and $\Gamma$ is the DCDM decay rate defined with respect to cosmic time.  These equations follow directly from covariant conservation of the stress-energy tensor for the combined DCDM+$\gamma$ system, and are analogous to those in models with DCDM decaying into dark radiation (DCDM-DR)~(e.g.,~\cite{Audren2014,Aoyama2014}).  Throughout, we assume a spatially flat universe.

The linear perturbation evolution equations for our model are similar to those of the DCDM-DR model, but some important differences arise due to the decay of the DCDM into Standard Model photons in our scenario.  We provide the perturbation equations of motion in the Supplemental Material.

We now briefly describe the parametrization of our model, as implemented in our modified version of the Einstein-Boltzmann solver {\tt CLASS}~\cite{Blas2011}.  We first define the temperature that the CMB photons would have at $z=0$ (today) \emph{if} no DCDM existed, $T_\mathrm{CMB,ini}$ (the subscript indicates that this quantity characterizes the photons at early times).  We then define $\delta T_{\rm CMB} \equiv T_\mathrm{CMB,0} - T_\mathrm{CMB,ini}$, the amount by which the CMB temperature today has been increased due to the DCDM, as compared to a no-DCDM universe.  Note that \emph{FIRAS} data require $T_\mathrm{CMB,0} \approx 2.7255$ K.  We then define the would-be density fraction in DCDM at $z=0$ if there had been no decay, $\Omega_{\rm DCDM, ini}$ (or $\omega_{\rm DCDM, ini} \equiv \Omega_{\rm DCDM, ini} h^2$),\footnote{As usual, $\Omega_i \equiv (\rho_{i,0}/\rho_{\mathrm{crit},0})$ is the energy density fraction in component $i$ at $z=0$ (today), where $\rho_{\mathrm{crit},0} = 3H_0^2/(8\pi G)$ is the critical density today ($G$ is Newton's constant), and $\omega_i \equiv \Omega_i h^2$, with $h \equiv H_0/(100 \, \mathrm{km/s/Mpc})$.} which sets the initial DCDM density in the early universe: $\rho_\mathrm{DCDM}(z_\mathrm{ini}) =  \Omega_\mathrm{DCDM,ini}\, \rho_{\mathrm{crit},0} \, (1+z_\mathrm{ini})^3$, with $1+z_{\rm ini} = 10^{14}$ ({\tt CLASS} default).  The initial density of decay photons also has an analytic form (proportional to $\Gamma$)~\cite{Audren2014}, although it is very close to zero for the models of interest.  We define $N_{\rm eff}$ with respect to $T_\mathrm{CMB,ini}$, as this is the relevant radiation energy density at high redshift: $\rho_{\mathrm{ur},0} = N_{\rm eff} (7/8) (4/11)^{4/3} (4 \sigma_\mathrm{B} T_\mathrm{CMB,ini}^4)$, where $\rho_{\mathrm{ur},0}$ is the energy density in ultra-relativistic (light) species today and $\sigma_\mathrm{B}$ is the Stefan-Boltzmann constant.\footnote{In practice, we sample in terms of the {\tt CLASS} parameter $N_{\rm ur}$ and treat $N_{\rm eff}$ as derived.}

\begin{table}
\begin{tabular}{c|c}
Parameter & Prior \tabularnewline
\hline 
$\hat{\omega}_\mathrm{b}$ & [0.005,0.1] \tabularnewline
$\hat{\omega}_\mathrm{c}$ & [0.001,0.99] \tabularnewline
$\ln(10^{10}\hat{A}_\mathrm{s})$ & [1.61,3.91] \tabularnewline
$n_s$ & [0.8,1.2] \tabularnewline
$100\theta_s$ & [0.5,10] \tabularnewline
$\tau_{\rm reio}$ & [0.01,0.8] \tabularnewline
$T_\mathrm{CMB,ini}/{\rm K}$ & [2.3,3.5] \tabularnewline
$\log_{10}(\hat{\omega}_\mathrm{DCDM,ini})$ & [-5,0] \tabularnewline
$\log_{10}(\Gamma/({\rm km/s/Mpc}))$ & [2,6] \tabularnewline
\hline 
$N_{\rm eff}$ & [0.01,10] \tabularnewline
\end{tabular}
\caption{DCDM$\gamma$ model parameters and uniform prior ranges used in MCMC analyses; we list $N_{\rm eff}$ separately as it is an extension of the base model.}\label{tab:priors}
\end{table}

Linear-theory CMB and LSS power spectra are nearly invariant under a certain rescaling of the cosmological parameters~\cite{Ivanov2020,Cyr-Racine:2021oal}, which motivates a reformulation of the fundamental cosmological parameter set in models with $T_\mathrm{CMB,0}$ allowed to vary.  Here we adapt the parameterization from Ref.~\cite{Ivanov2020} to our scenario:
\ba
    \hat{\omega}_\mathrm{b} & \equiv & \omega_\mathrm{b} \left(\frac{T_\mathrm{CMB,ini}}{T_{_\mathrm{FIRAS}}}\right)^{-3} \\
    \hat{\omega}_\mathrm{c} & \equiv & \omega_\mathrm{c} \left(\frac{T_\mathrm{CMB,ini}}{T_{_\mathrm{FIRAS}}}\right)^{-3} \\
    \hat{\omega}_\mathrm{DCDM,ini} & \equiv & \omega_\mathrm{DCDM,ini} \left(\frac{T_\mathrm{CMB,ini}}{T_{_\mathrm{FIRAS}}}\right)^{-3} \\
    \hat{A}_\mathrm{s} & \equiv & A_\mathrm{s} \left(\frac{T_\mathrm{CMB,ini}}{T_{_\mathrm{FIRAS}}}\right)^{n_\mathrm{s}-1} \,.
\ea
with $T_{_\mathrm{FIRAS}}=2.7255 \, \mathrm{K}$. Sampling the posterior in these parameters is more efficient~\cite{Ivanov2020}.  The complete set of free parameters in our model and the priors used in our Markov Chain Monte Carlo (MCMC) analyses are given in Table~\ref{tab:priors}.  We adopt uniform priors on $\log_{10}(\hat{\omega}_\mathrm{DCDM,ini})$, as this quantity is expected to be small, and on $\log_{10}(\Gamma)$, to explore a wide range of decay redshifts.  We have validated our implementation against that of Ref.~\cite{Ivanov2020} for models with $T_{\rm CMB,0} \neq 2.7255$~K (but with $T_{\rm CMB}(z) \propto (1+z)$), to ensure that subtleties related to Big Bang nucleosynthesis (BBN) and CMB modeling (see~\cite{Ivanov2020}) are treated properly.

Following \emph{Planck}~\cite{Planck2018parameters}, we fix the sum of the neutrino massses to $\sum m_{\nu} = 0.06$ eV, with one massive eigenstate and two massless eigenstates.  The primordial helium fraction is computed via BBN.  Nonlinear effects in the matter power spectrum are calculated via {\tt Halofit}~\cite{Smith:2002dz,Takahashi2012}, although the observables we consider are dominated by linear modes.
%%%%%%%%%%%%%%%%%%%%%%%%%%%%%%%%%%%%%%%%%%%%%%

%%%%%%%%%%%%%%%%%%%%%%%%%%%%%%%%%%%%%%%%%%%%%%
\emph{Phenomenology---}
In the DCDM$\gamma$ model, a natural expectation is that the redshift of recombination, $z_\star$, can be increased by lowering $T_{\rm CMB,ini}$, thus yielding a smaller sound horizon, $r_s^\star$.  However, the decrease in the radiation density due to the lower $T_{\rm CMB,ini}$, and hence the decrease in $H(z)$ in the early universe, actually leads to an \emph{increase} in $r_s^\star$.  To keep $\theta_s \equiv r_s^\star/D_A^\star$ fixed, the angular diameter distance to recombination, $D_A^\star = \int _0^{z_\star} c \, {\rm d} z/H(z)$, must also increase.  However, due to the increase in $z_\star$, the integral appearing in $D_A^\star$ actually increases by too much, and one must \emph{increase} $H_0$ to compensate for this and leave $\theta_s$ unchanged.  In addition, to keep $k_{\rm eq}$ (the scale of matter-radiation equality) fixed, the matter density must decrease, which decreases $S_8$ (the small amount of dark matter that decays also helps in this regard).  When fit to CMB data, the PRR scenario thus naturally moves both $H_0$ and $S_8$ toward values that relax tensions with direct probes;\footnote{The background physics underlying the increase in $H_0$ is identical to that in $\Lambda$CDM with varying $T_{\rm CMB,0}$~\cite{Ivanov2020}; the changes in $H(z)$ due to PRR are very small, and its primary role is to match \emph{FIRAS}.  Furthermore, the novel terms in the perturbation evolution equations in DCDM$\gamma$ play an almost negligible role in the CMB fit (see BH23).} an increase $\delta T_{\rm CMB} \approx 100$ mK in the post-recombination epoch would suffice to resolve the discrepancies.

Generically, achieving this phenomenology requires the DCDM to decay after recombination.  Due to observational constraints requiring $T_{\rm CMB}(z) \propto (1+z)$ at $z \lesssim 3$~(e.g.,~\cite{Saro2014,Luzzi2015,Avgoustidis2016,Li_ACT_2021}), we thus expect the DCDM to decay during matter domination.\footnote{A constraint on $T_{\rm CMB}$ at $z=6.34$ was recently reported~\cite{Riechers2022}; however, the error bar is sufficiently large to encompass the small temperature change in the model proposed here, and we are primarily focused on changes to $T_{\rm CMB}$ at higher redshifts.}  During this epoch, the DCDM decay redshift, $z_X$, is well-approximated by \cite{bcb2021} 
%$z_X \approx [(\Gamma/(95.6 \, {\rm km/s/Mpc}))((70 \, {\rm km/s/Mpc})/H_0)\sqrt{0.3/\Omega_\mathrm{m}}]^{2/3}$; 
$z_X \approx 0.812 \, (\Gamma/H_0)^{2/3} (0.3/\Omega_{\rm m})^{1/3}$; 
e.g., $\Gamma = 10^4$ km/s/Mpc yields $z_X \approx 22$.  Our prior range for $\Gamma$ (c.f.~Table~\ref{tab:priors}) thus corresponds to $1 \lesssim z_X \lesssim 500$. A crucial feature of decay during matter domination is that even a small fraction of DCDM can inject enough energy to change $T_{\rm CMB}$ significantly, since $\rho_{\rm c}(z) \gg \rho_\gamma(z)$ during this epoch.  For example, at $z_X = 22$, $\rho_{\rm c} \approx 210 \, \rho_{\gamma}$, and thus increasing $\rho_\gamma$ by 4\% at this epoch (i.e., increasing $T_{\rm CMB}(z=22)$ by 1\%) requires only $\simeq 0.02$\% of the CDM to decay.  Thus, our prior range on the DCDM density extends to very small values (c.f.~Table~\ref{tab:priors}), motivating the use of a uniform logarithmic prior.

A crucial assumption underlying our model is that the photon distribution function maintains its blackbody form to guarantee consistency with \emph{FIRAS} data, which severely constrain CMB monopole spectral distortions~\cite{FIRAS1996,bcb2021,Bianchini:2022dqh}.  We assume that the injected energy is fully thermal and simply increases the monopole CMB temperature.  If photons are injected near the peak of the CMB monopole ($\approx 160$ GHz today), \emph{FIRAS} constrains the relative amount of decaying DM, $f_\mathrm{DCDM} \equiv {\omega}_\mathrm{DCDM,ini}/{\omega}_\mathrm{c,ini}$, to $f_\mathrm{DCDM} \lesssim 10^{-4}$ \cite{bcb2021}. Nonetheless, if the injection happens at frequencies a factor of a few lower than or a factor of 100 higher than the peak, $f_\mathrm{DCDM}$ is not constrained because for such photons the universe is transparent and no spectral distortion is generated in the \emph{FIRAS} frequency range (see~\cite{bcb2021} and BH23 for further details).  However, one would need to track the full set of spatial-spectral equations in such models~\cite{Chluba:2022xsd,Chluba:2022efq,Kite:2022eye}, as the radiation would no longer be blackbody.  Here we adopt the simpler approach of assuming that the injected radiation is thermal; at the background level, particularly for $H_0$, this assumption makes no difference.  Although it is challenging to construct models that fully thermalize extra radiation in the post-recombination universe \cite{Chluba:2014wda}, injecting photons outside the range constrained by \emph{FIRAS} is straightforward.  Here we assume that linear perturbation theory in such models is well-approximated by our approach, given that the injected energy and $f_\mathrm{DCDM}$ are quite small.  This assumption will be weakly violated in a full particle physics model, but corrections to the above equations will necessarily be small if the model is to satisfy the \emph{FIRAS} constraints.

BBN abundance constraints are another potential challenge.  However, we demonstrate in the Supplemental Material that the DCDM$\gamma$ and DCDM$\gamma$+$N_{\rm eff}$ models both fit BBN data well.
%%%%%%%%%%%%%%%%%%%%%%%%%%%%%%%%%%%%%%%%%%%%%%

%%%%%%%%%%%%%%%%%%%%%%%%%%%%%%%%%%%%%%%%%%%%%%
\emph{Data---}
We constrain the DCDM$\gamma$(+$N_{\rm eff}$) parameters using the following datasets:
\begin{itemize}
    \item {\it FIRAS} $T_{\rm CMB,0}$: Fixsen (2009)~\cite{Fixsen_2009} measurement of $T_{\rm CMB,0} = 2.72548 \pm 0.00057$ K;
    \item {\it CMB TT+TE+EE}: \emph{Planck} PR3 (2018) primary CMB TT, TE, and EE power spectra~\cite{Planck2018likelihood,Planck2018overview,Planck2018parameters};
    \item {\it CMB Lensing $\phi\phi$}: \emph{Planck} PR3 reconstructed CMB lensing potential power spectrum~\cite{Planck2018lensing};
    \item {\it BAO}: Baryon acoustic oscillation (BAO) distances from
    SDSS~\cite{Ross:2014qpa,Alam:2016hwk} and 6dF~\cite{2011MNRAS.416.3017B};
    \item {\it SNIa}: Pantheon Type Ia supernova (SNIa) distances~\cite{Scolnic:2017caz}; 
    \item {\it RSD}: Redshift-space distortion (RSD) constraints on $f\sigma_8(z)$ from SDSS~\cite{Satpathy:2016tct,Alam:2016hwk} (we include the BAO-RSD covariance);
    \item {\it DES-Y3}: DES-Y3~\cite{DES:2021wwk} measurement of $S_8 = 0.776 \pm 0.017$ (we verify in BH23 that the full DES ``3$\times$2pt'' likelihood is well-approximated by this prior on $S_8$ using the same approach as in~\cite{Hill:2020osr});
    \item {\it SH0ES}: Direct constraint on $H_0$ from Cepheid-calibrated SNIa ($H_0 = 73.2 \pm 1.3$ km/s/Mpc)~\cite{Riess:2020fzl};
    \item {\it Combined} $H_0$: Inverse-variance-weighted combination of direct constraints on $H_0$ from three independent probes: SH0ES, TRGB-calibrated surface brightness fluctuations translated into the 2M++ frame ($H_0 = 73.6 \pm 3.6$ km/s/Mpc)~\cite{Blakeslee:2021rqi}, and megamaser distances with peculiar velocity corrections in the 2M++ frame ($H_0 = 70.1 \pm 2.9$ km/s/Mpc)~\cite{Boruah:2020fhl}, yielding $H_0 = 72.8 \pm 1.1$ km/s/Mpc.\footnote{Our results would be nearly unchanged if we instead used the most recent SH0ES measurement on its own, $H_0 = 73.04 \pm 1.04$ km/s/Mpc~\cite{Riess:2021jrx}.  However, the mild preference that we find for the DCDM$\gamma$ model would be weakened if we instead used $H_0 = 69.8 \pm 0.6 \, ({\rm stat.}) \pm 1.6 \, ({\rm syst.})$ km/s/Mpc from TRGB-calibrated SNIa~\cite{Freedman:2021ahq}.}
\end{itemize}

We sample the parameter posteriors using the MCMC code {\tt Cobaya}~\cite{Torrado:2020dgo} coupled to our modified {\tt CLASS}.  We consider our MCMC chains converged when the Gelman-Rubin~\cite{Gelman:1992zz} criterion $R-1 < 0.03$.
%%%%%%%%%%%%%%%%%%%%%%%%%%%%%%%%%%%%%%%%%%%%%%

%%%%%%%%%%%%%%%%%%%%%%%%%%%%%%%%%%%%%%%%%%%%%%
\emph{Results---}
The main results of our analysis are presented in Fig.~\ref{fig:DCDMgamma}.  In the Supplemental Material, we include full numerical constraints (Table~\ref{table:DCDMgamma}) and results for DCDM$\gamma$+$N_{\rm eff}$ (Table~\ref{table:DCDMgammaNeff} and Fig.~\ref{fig:DCDMgammaNeff}).\footnote{We also explore an extension with varying curvature density $\Omega_k$, but find that the DCDM$\gamma$ constraints are generally within $1\sigma$ of those in the $\Omega_k = 0$ case.}  For DCDM$\gamma$, primary CMB data alone show no preference for the model over $\Lambda$CDM.\footnote{Marginalization effects~(e.g.,~\cite{Gomez-Valent:2022hkb}) associated with the prior volume of $\Gamma$ in the $\omega_{\rm DCDM} \rightarrow 0$ limit could mildly bias constraints toward $\Lambda$CDM in this analysis.}  However, the inclusion of SH0ES $H_0$ data leads to a mild preference for non-zero DCDM: $\log_{10}(\hat{\omega}_\mathrm{DCDM,ini}) = -3.77^{+0.66}_{-1.1}$, and an accordingly higher $H_0 = 71.2 \pm 1.1$ km/s/Mpc and lower $S_8 = 0.774 \pm 0.018$ with $\delta T_{\rm CMB} = 0.109^{+0.033}_{-0.044}$~K.  The fit to \emph{Planck} CMB data is slightly degraded, with $\Delta \chi^2_{Planck} = 6.7$ compared to the best-fit $\Lambda$CDM model to \emph{Planck} alone.  Removing SH0ES but including the full suite of non-$H_0$ datasets (``walking barefoot''~\cite{Hill:2020osr}), the preference for DCDM weakens but is somewhat stronger than for primary CMB data alone, due to the influence of the DES-Y3 $S_8$ constraint.  Using all datasets, including all three independent $H_0$ probes, we find $H_0 = 68.69 \pm 0.35$ km/s/Mpc, $S_8 = 0.8035 \pm 0.0081$, and $\delta T_{\rm CMB} < 0.0342$~K (95\% CL upper limit).  Interestingly, the fit to \emph{Planck} is not degraded compared to that of $\Lambda$CDM, with $\Delta \chi_{Planck}^2 = 0.3$, even though $H_0$ increases by $2.5\sigma$ from its $\Lambda$CDM value ($67.36 \pm 0.54$ km/s/Mpc)~\cite{Planck2018parameters}.

\begin{figure*}[!tp]
\includegraphics[width=\textwidth]{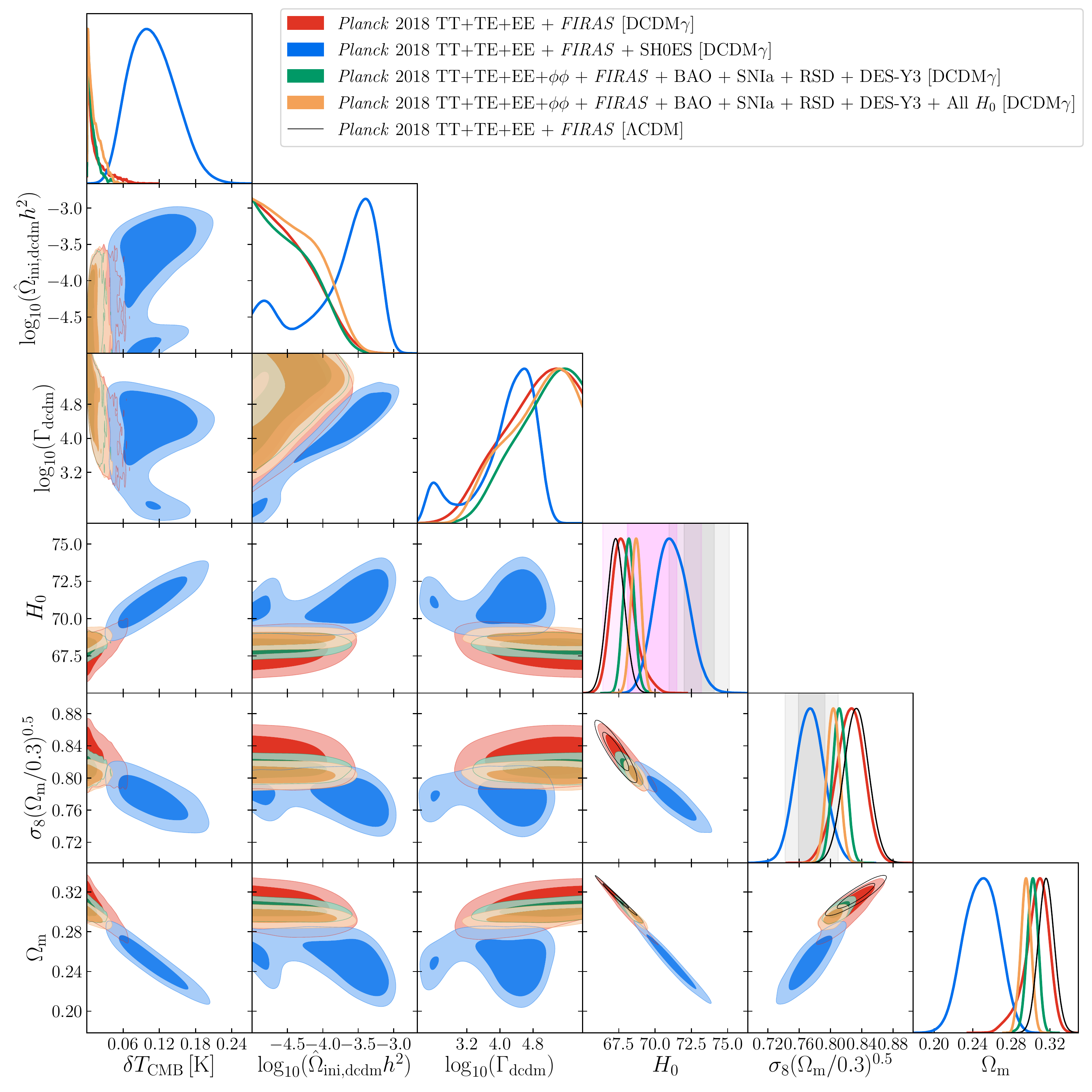}
\caption{Marginalized posteriors for the DCDM$\gamma$ parameters and select other parameters in fits to various datasets, as labeled, with $H_0$ and $\Gamma$ in units of km/s/Mpc.  For comparison, $\Lambda$CDM constraints for \emph{Planck} CMB and \emph{FIRAS} data are shown in thin black lines.  The vertical grey and magenta bands in the $H_0$ panel show the latest SH0ES~\cite{Riess:2021jrx} and TRGB~\cite{Freedman:2021ahq} measurements, respectively.  The vertical grey band in the $S_8$ panel shows the DES-Y3 constraint~\cite{DES:2021wwk}.}
\label{fig:DCDMgamma}
\end{figure*}

Although the DCDM component is not strongly detected, all dataset combinations exclude low values of $\Gamma$ (i.e., low $z_X$).  In the \emph{Planck}+\emph{FIRAS}+SH0ES analysis, the best-fit $z_X \simeq 40$; in the joint analysis of all datasets, the best-fit $z_X \simeq 190$ (and $z_X \gtrsim 10$ at 95\% CL).  The (mildly) preferred decay redshifts are thus prior to reionization, where no direct observational constraints exist.

A key takeaway is that a small amount of DCDM can significantly affect the radiation energy density during this epoch.  In the \emph{Planck}+\emph{FIRAS}+SH0ES analysis, the best-fit $f_{\rm DCDM} = 0.332$\%, yielding $\delta T_{\rm CMB} = 0.138$ K; in the joint analysis of all datasets, the best-fit $f_{\rm DCDM} = 0.153$\%, yielding $\delta T_{\rm CMB} = 0.0169$ K.  Notably, our upper bounds on $f_{\rm DCDM}$ are an order of magnitude below those on the energy density of massive neutrinos, $\omega_\nu / \omega_c \lesssim 1.1$\%~\cite{Planck2018parameters}; thus, any effects on early structure formation due to the potentially large thermal velocities of the DCDM component are negligible.

The primary obstacle to further increasing $H_0$ in the global fit to DCDM$\gamma$ comes from the independent constraint on $\Omega_{\rm m}$ from the BAO and SNIa data, which do not allow $\Omega_{\rm m}$ to take the low value obtained in the fit to \emph{Planck}+\emph{FIRAS}+SH0ES.  This low value of $\Omega_{\rm m}$ is needed to maintain $k_{\rm eq}$ in the fit to \emph{Planck}, due to the decreased radiation density at early times.  A simple remedy to this problem is to introduce additional relativistic energy density in the early universe via $N_{\rm eff} > 3.044$.  Results of this analysis are reported in the Supplemental Material.

%%%%%%%%%%%%%%%%%%%%%%%%%%%%%%%%%%%%%%%%%%%%%%
\emph{Outlook---}
In this work, we have investigated the extent to which cosmological data permit PRR, even if the injected photons preserve the CMB blackbody form.  In the DCDM$\gamma$ model, we find tight limits, with $\delta T_{\rm CMB} < 0.0342$~K (95\% CL, all datasets).  These constraints are fully independent from CMB spectral distortion probes, and similarly tight bounds should hold for other specific PRR models.  Such a reheating episode could play a role in resolving current tensions between inferences of $H_0$ and $S_8$, although constraints on $\Omega_{\rm m}$ obstruct a complete resolution without further modifications.

The PRR impact on $H_0$ differs from the modified recombination scenario of Ref.~\cite{Chiang:2018xpn}, who suggest a modification to the CMB spectrum deep in the Wien tail, where \emph{FIRAS} constraints are absent ($\gtrsim 1.5$ THz); such a model does not change $\rho_{\gamma}(z)$ significantly, instead only removing a small fraction of the high-energy photons to change the timing of recombination.  Our scenario also differs from modified-recombination models that increase $H_0$ via a decrease in $r_s^\star$, such as via primordial magnetic fields~(e.g.,~\cite{Jedamzik:2020krr, Thiele:2021okz}) or an increase in the electron mass~(e.g.,~\cite{Sekiguchi:2020teg, Hart:2019dxi, Lee:2022gzh}).  In contrast, $r_s^\star$ (slightly) increases in the PRR scenario, and $H_0$ is increased due to the interplay with $D_A^\star$ discussed earlier.

Tighter PRR constraints can be obtained using CMB data from the Atacama Cosmology Telescope~\cite{Choi2020,Aiola2020} and galaxy clustering data from the Baryon Oscillation Spectroscopic Survey (BOSS)~\cite{Philcox:2020vvt}, amongst other probes~\cite{Surrao2023}.  Moreover, the difference in the spectral energy distribution of the CMB anisotropies in PRR scenarios compared to that of the monopole at $z = 0$ leads to spatial-spectral distortions, which can be constrained via the formalism of Refs.~\cite{Chluba:2022xsd,Chluba:2022efq,Kite:2022eye}; this approach is formally necessary for models in which the DCDM produces photons in a line outside the \emph{FIRAS} frequency range (thus increasing $\rho_\gamma$, but in a non-thermal sub-component).

Many models have been constructed in which dark states decay through various channels that eventually cascade to photons~(e.g.,~\cite{Chen2004,Zhang2007,Ruderman2010,Ackermann2015,Ando2015,Becker2020,Jaeckel:2021ert}), or in which metastable vacuum decay produces additional photons~\citep{Freese1987,Overduin1993,Lima1996,Opher2005}.  Ref.~\cite{Sobotka:2022vrr} considered a scenario similar to ours, but with decays prior to recombination rather than afterward, finding tight bounds solely from background cosmological signatures.  Models with heavier DM decaying to higher-energy photons ($\gtrsim$ keV) have been constrained via their impact on the ionization and thermal history of the universe, searches for direct signatures in X-ray and $\gamma$-ray data, and other probes~(e.g.,~\cite{Chen2004,Padmanabhan:2005es,Galli:2009zc,Madhavacheril:2013cna,Slatyer:2016qyl,Green:2018pmd,Planck2018parameters,Liu:2023fgu,Liu:2023nct}); we have considered a scenario in which such bounds are negligible due to the low energy of the decay photons.

Small modifications of the DCDM$\gamma$ scenario considered here may be necessary to evade bounds from stellar cooling and supernovae (e.g.,~\cite{Raffelt:2006cw,Isern2008,Ayala2014}).  This can be achieved by replacing the particle decay process with radiation arising from the decay of an excited state in the dark sector to its ground state (e.g.,~\cite{Tucker-Smith:2001myb,Alexander:2016aln,Baryakhtar:2020rwy,CarrilloGonzalez:2021lxm}).  In this model, the DM possesses a dipole moment and couples to the Standard Model through a kinetically mixed massive dark photon, which allows for transitions between the ground and excited DM states.  For an energy splitting $\simeq 0.1$ eV between these states, and theoretically reasonable transition dipole moments, the model yields lifetimes $\simeq 1-10$ Myr (see Eq.~2.2 of~\cite{Baryakhtar:2020rwy}), which is exactly the regime of interest for the DCDM$\gamma$ scenario.  If the massive dark photon is heavier than an MeV, it is not produced in stars or supernovae, and bounds on interactions today are irrelevant since essentially all of the energy density will have decayed away by the present time. The primary change to the model considered here is the treatment of the DCDM density, but this is already constrained to be sufficiently small that we expect our implementation to yield similar results to this approach, which we leave to future work.

Our results indicate that, independent of spectral distortion constraints, cosmological data allow only small changes to the photon energy density between recombination and the present era.  Fundamentally, this arises from the agreement between the \emph{FIRAS} direct measurement of $T_{\rm CMB,0}$~\cite{Fixsen_2009} and the indirect inference from CMB, BAO, and other data~\cite{Ivanov2020}.  Future high-precision CMB spectrometers can further constrain PRR models~\cite{Kogut2011,Maffei:2021xur,Chluba:2019kpb,Chluba:2019nxa,Kogut:2019vqh}.
%%%%%%%%%%%%%%%%%%%%%%%%%%%%%%%%%%%%%%%%%%%%%%

%TC:ignore
%%%%%%%%%%%%%%%%%%%%%%%%%%%%%%%%%%%%%%%%%%%%%%
{\small \emph{Acknowledgments.} We are grateful to Ameya Chavda, Fiona McCarthy, Adam Riess, and Ken van Tilburg for useful exchanges, and especially to Mikhail Ivanov for helpful comments and providing code from Ref.~\cite{Ivanov2020} for validation.  JCH acknowledges support from NSF grant AST-2108536, NASA grants 21-ATP21-0129 and 22-ADAP22-0145, DOE grant DE-SC00233966, the Sloan Foundation, and the Simons Foundation.  We thank the Scientific Computing Core staff at the Flatiron Institute for computational support.  The Flatiron Institute is supported by the Simons Foundation.  BB acknowledges  support from the European Research Council (ERC) under the European Union’s Horizon 2020 research and innovation programme (Grant agreement No. 851274). We acknowledge use of the {\tt matplotlib}~\cite{Hunter2007}, {\tt numpy}~\cite{2020Natur.585..357H}, {\tt GetDist}~\cite{Lewis:2019xzd},\footnote{\url{https://github.com/cmbant/getdist}} and {\tt Cobaya}~\cite{Torrado:2020dgo}\footnote{\url{https://github.com/CobayaSampler/cobaya}} packages, use of the Boltzmann code {\tt CLASS}~\cite{Blas2011},\footnote{\url{http://class-code.net/}} and use of the BBN code {\tt PRIMAT}~\cite{Pitrou:2018cgg,Pitrou:2020etk}.\footnote{\url{http://www2.iap.fr/users/pitrou/primat.htm}}}
%%%%%%%%%%%%%%%%%%%%%%%%%%%%%%%%%%%%%%%%%%%%%%

\clearpage

%%%%%%%%%%%%%%%%%%%%%%%%%%%%%%%%%%%%%%%%%%%%%%
\section*{Supplemental Material}

\subsection*{Linear Perturbation Evolution Equations}

For component $i$, defining the density perturbation $\delta_i \equiv \rho_i/\bar{\rho}_i - 1$ and the velocity divergence $\theta_i \equiv i k^j v_j$ where $k^j$ is the Fourier-space wavevector (we follow the definitions and notation of~\cite{MaBertschinger1995}), the evolution of the DCDM and photon perturbations are, respectively:
\ba
    \delta_\mathrm{DCDM}^\prime & = & -\theta_\mathrm{DCDM} - m_\mathrm{cont} - a\Gamma m_\psi \label{eq:ddcdm} \\
    \theta_\mathrm{DCDM}^\prime & = & -\frac{a^\prime}{a}\theta_\mathrm{DCDM} + k^2m_\psi \label{eq:tdcdm} \\
    \delta_{\gamma}^\prime & = & -\frac{4}{3}\theta_{\gamma} - \frac{4}{3}m_\mathrm{cont} \nonumber \\
    						& & + a\Gamma\frac{\rho_\mathrm{DCDM}}{\rho_{\gamma}}\left(\delta_\mathrm{DCDM} - \delta_{\gamma} + m_{\psi}\right) \label{eq:EoMgdra} \\
    \theta_{\gamma}^\prime & = & k^{2}\left(\frac{1}{4}\delta_{\gamma} - \sigma_{\gamma}\right) + k^{2} m_{\psi} + an_{e}\sigma_\mathrm{T}\left(\theta_{b} - \theta_{\gamma}\right) \nonumber \\ 
    						& & - \frac{3}{4}a\Gamma\frac{\rho_\mathrm{DCDM}}{\rho_{\gamma}}\left(\frac{4}{3}\theta_{\gamma} - \theta_\mathrm{DCDM}\right) \label{eq:EoMgdrb}\\
    F_{\gamma,2}^\prime & = & 2\sigma_{\gamma}^\prime=\frac{8}{15}\theta_{\gamma}-\frac{3k}{5}F_{\gamma,3} + \frac{8}{15}m_\mathrm{shear} \nonumber \\
    						& & - \frac{9}{5}an_{e}\sigma_\mathrm{T}\sigma_{\gamma}+\frac{1}{10}an_{e}\sigma_\mathrm{T}\left(G_{\gamma,0}+G_{\gamma,2}\right) \nonumber \\
						& & - 2\sigma_\gamma a\Gamma\frac{\rho_\mathrm{DCDM}}{\rho_{\gamma}} \label{eq:EoMgdrc}\\
    F_{\gamma,\ell}^\prime & = & \frac{k}{2\ell+1}\left[\ell F_{\gamma,\ell-1}-\left(\ell+1\right)F_{\gamma,\ell+1}\right] \nonumber \\
    						& & - a\Gamma F_{\gamma,\ell}\frac{\rho_\mathrm{DCDM}}{\rho_{\gamma}} \label{eq:EoMgdrd}
\ea
where the metric terms ($m_{\rm cont}$, $m_\psi$, $m_{\rm shear}$) depend on the gauge as written in Table~\ref{tab:mc}, $\sigma_{\rm T}$ is the Thomson cross-section, $n_e$ is the electron density, $G_\gamma$ is the difference between the two linear polarization components~\cite{MaBertschinger1995}, and $F_\gamma$ is the integrated perturbed phase-space distribution~\cite{Kaplinghat1999}:
\be
    F_\gamma = \frac{\int \mathrm{d}q \, q^3 \, f_\gamma^0 \, \Psi_\gamma}{\int \mathrm{d}q \, q^3 \, f_\gamma^0} \,, 
    \label{eq:Fdr}
\ee
with $F_{\gamma,\ell}$ the Legendre series expansion coefficients of this first-order quantity.  In Eq.~\eqref{eq:Fdr}, $q$ is the comoving photon spatial momentum and $f_\gamma^0 \equiv f_\gamma^0(q,\tau)$ is the unperturbed (background) part of the phase-space distribution, i.e., $f_\gamma(\vec{x}, \vec{q}, \tau) = f_\gamma^0(q,\tau) (1 + \Psi_\gamma(\vec{x}, \vec{q}, \tau))$, which defines the perturbation $\Psi_\gamma$.  In the photon perturbation evolution equations, the new terms introduced in our model are those on the RHS of Eqs.~\eqref{eq:EoMgdra}--\eqref{eq:EoMgdrd} that are proportional to $\Gamma$; the others are identical to those in the standard model~\cite{MaBertschinger1995}.  The photon polarization evolution equations remain the same as in $\Lambda$CDM.

\begin{table}
%\begin{centering}
\begin{tabular}{c|c|c}
%\multicolumn{1}{c}{Gauge} & Synchronous & \multicolumn{1}{c}{Newtonian}\tabularnewline
Gauge & Synchronous & Newtonian \tabularnewline
\hline 
$m_{\mathrm{cont}}$ & $ h^\prime/2$ & $-3 \phi^\prime $\tabularnewline
\hline 
$m_{\psi}$ & $0$ & $\psi$\tabularnewline
\hline 
$m_{\mathrm{shear}}$ & $\left(h^\prime + 6 \eta^\prime \right)/2$ & $0$\tabularnewline
\end{tabular}
%\par\end{centering}
\caption{Metric terms appearing in the linear perturbation theory equations for our model, in the synchronous and Newtonian gauges ($(h,\eta)$ and $(\phi,\psi)$ are the metric perturbations in these gauges; see Ref.~\cite{MaBertschinger1995} for definitions).}\label{tab:mc}
\end{table}

\subsection*{BBN Constraints}

Due to the lower radiation density at early times, one might expect that BBN constraints could present a challenge to our scenario.  Fortunately, the light element yields are predominantly controlled by $\hat{\omega}_\mathrm{b}$, which is also the parameter combination involving $\omega_b$ and $T_{\rm CMB,ini}$ to which the CMB is most sensitive~\cite{Ivanov2020}.  Thus, our best-fit models yield BBN results that hardly differ from those of the best-fit $\Lambda$CDM model.  We explicitly verify this result using the BBN code {\tt PRIMAT}~\cite{Pitrou:2018cgg,Pitrou:2020etk}.  For the best-fit DCDM$\gamma$ (DCDM$\gamma$+$N_{\rm eff}$) model to the full set of cosmological data, we find the He abundance $Y_{\rm P} = 0.24720$ and D abundance $n_{\rm D}/n_{\rm H} = 2.444 \times 10^{-5}$ ($Y_{\rm P} = 0.25074$ and $n_{\rm D}/n_{\rm H} = 2.480 \times 10^{-5}$), while for the best-fit $\Lambda$CDM model to \emph{Planck} CMB data we have $Y_{\rm P} = 0.24721$ and $n_{\rm D}/n_{\rm H} = 2.438 \times 10^{-5}$.  Recent observational data yield $Y_{\rm P} = 0.2449 \pm 0.0040$~\cite{Aver:2015iza} or $Y_{\rm P} = 0.2446 \pm 0.0029$~\cite{Peimbert:2016bdg}, and $n_{\rm D}/n_{\rm H} = 2.527 \pm 0.030 \times 10^{-5}$~\cite{Cooke:2017cwo}.  Our results are consistent with these measurements; in fact the DCDM$\gamma$ and DCDM$\gamma$+$N_{\rm eff}$ models improve the agreement with the D abundance data.

\subsection*{DCDM$\gamma$ Results}

A full compilation of the marginalized constraints on the DCDM$\gamma$ model parameters is provided in Table~\ref{table:DCDMgamma}.

\begin{table*}[ht!]
Constraints on DCDM$\gamma$ Parameters  \\
  \centering
  %\begin{tabular}{|c|c|c|c|c|c|}
  \begin{tabular}{|c|c|c|c|c|}
    \hline\hline Parameter & \hspace{0cm} 
    \begin{tabular}[t]{@{}c@{}}\emph{Planck} CMB \\ TT+TE+EE, \\ \emph{FIRAS} \end{tabular} & \hspace{0cm}
    \begin{tabular}[t]{@{}c@{}}\emph{Planck} CMB \\ TT+TE+EE, \\ \emph{FIRAS}, SH0ES \end{tabular} & \hspace{0cm}
    %\begin{tabular}[t]{@{}c@{}}\emph{Planck} CMB\\ TT+TE+EE, \\  \emph{FIRAS}, BAO \end{tabular} & \hspace{0cm}
    \begin{tabular}[t]{@{}c@{}}\emph{Planck} CMB\\ TT+TE+EE, \\  \emph{FIRAS}, BAO, \\ \emph{Planck} $\phi\phi$, SNIa, \\ RSD, DES-Y3\end{tabular} &
    \hspace{0cm}
    \begin{tabular}[t]{@{}c@{}}\emph{Planck} CMB\\ TT+TE+EE, \\  \emph{FIRAS}, BAO, \\ \emph{Planck} $\phi\phi$, SNIa, \\ RSD, DES-Y3, \\ Combined $H_0$\end{tabular}
    \\\hline \hline

{$T_\mathrm{CMB,ini}$} [K] & $2.710 (2.718)^{+0.016}_{-0.0024}  $  &  $2.617 (2.587)^{+0.044}_{-0.033}$  &  $2.7158 (2.7230)^{+0.0093}_{-0.0023}$   & $2.712 (2.709)^{+0.013}_{-0.0044}$    \\

{$\log_{10}(\hat{\omega}_\mathrm{DCDM,ini})$} &  $< -3.78 (-4.35)$ & $-3.77 (-3.47)^{+0.66}_{-1.1}$    &   $< -3.82 (-4.44)$   & $< -3.74 (-3.75)$   \\

{$\log_{10}(\frac{\Gamma}{{\rm km/s/Mpc}})$} & $> 3.36 (5.01)$  & $4.12 (4.36)^{+0.89}_{-0.26}$ & $> 3.75 (5.67)$     & $> 3.49 (5.40)$   \\

{$\ln(10^{10}\hat{A}_\mathrm{s})$} &$3.044 (3.050) \pm 0.016$  &  $3.043 (3.034) \pm 0.016$   & $3.039 (3.049)\pm 0.014$   & $3.045 (3.039) \pm 0.013$    \\

{$n_\mathrm{s}   $} & $0.9647 (0.9654) \pm 0.0043$  & $0.9677 (0.9677) \pm 0.0043$  &   $0.9671 (0.9697) \pm 0.0037$   & $0.9691 (0.9679) \pm 0.0036$   \\

% formatting compression
%{$100\theta_\mathrm{s}$} & $1.04186 (1.04181) \pm 0.00030$ & $1.04197 (1.04210) \pm 0.00030$   & $1.04193 (1.04201) \pm 0.00028$    & $1.04204 (1.04193) \pm 0.00028$ \\

{$100\theta_\mathrm{s}$} $\times 10^4$ & $10418.6 (10418.1) \pm 3.0$ & $10419.7 (10421.0) \pm 3.0$   & $10419.3 (10420.1) \pm 2.8$  & $10420.4 (10419.3) \pm 2.8$ \\

% formatting compression
%{$\hat{\Omega}_\mathrm{b} h^2$} &$0.02234 (0.02237) \pm 0.00015$  &  $0.02242 (0.02230) \pm 0.00015$     &   $0.02243 (0.02250) \pm 0.00013$    & $0.02252 (0.02237) \pm 0.00014$   \\

{$\hat{\Omega}_\mathrm{b} h^2$} $\times 100$ & $2.234 (2.237) \pm 0.015$  &  $2.242 (2.230) \pm 0.015$     &   $2.243 (2.250) \pm 0.013$    & $2.252 (2.237) \pm 0.014$   \\

{$\hat{\Omega}_\mathrm{c} h^2$} & $0.1201 (0.1200) \pm 0.0014$  & $0.1190 (0.1192) \pm 0.0014$  &   $0.11878 (0.11822) \pm 0.00092$   & $0.11803 (0.11859)^{+0.00083}_{-0.00099}$   \\

{$\tau_\mathrm{reio}$} & $0.0541 (0.0556) \pm 0.0076$  &  $0.0542 (0.0481) \pm 0.0080$  &   $0.0531 (0.0576) \pm 0.0069$    & $0.0562 (0.0536) \pm 0.0069$    \\

    \hline

$\delta T_{\rm CMB}$ [K]   &  $< 0.0535 (0.0077)$  & $0.109 (0.138) ^{+0.033}_{-0.044}$  &   $< 0.0279 (0.0024)$   & $< 0.0342 (0.0169) $ \\

$H_0       $ [km/s/Mpc]      &$67.82 (67.59)^{+0.63}_{-0.88}$  & $71.2 (71.9) \pm 1.1$ &  $68.19 (68.26) \pm 0.39$     & $68.69 (68.48) \pm 0.35$\\

$\Omega_\mathrm{c} h^2     $ &  $0.1180 (0.1190)^{+0.0029}_{-0.0014}$  & $0.1054 (0.1020)^{+0.0047}_{-0.0040}$  &   $0.1175 (0.1179)^{+0.0012}_{-0.00092}$   & $0.1163 (0.1164)^{+0.0013}_{-0.00095}$\\

$\Omega_\mathrm{m}         $ & $0.306 (0.311)^{+0.015}_{-0.0090}  $   & $0.249 (0.235) \pm 0.017$ &  $0.3019 (0.3026)^{+0.0059}_{-0.0053}$     & $0.2950 (0.2963)^{+0.0056}_{-0.0050}$\\

$\sigma_8                  $ & $0.8173 (0.8164)^{+0.0085}_{-0.011} $  & $0.850 (0.860) ^{+0.015}_{-0.020}$   &   $0.8089 (0.8087)^{+0.0054}_{-0.0071}$     & $0.8104 (0.8120)^{+0.0064}_{-0.0078}$\\

$S_8$                        & $0.825 (0.831)^{+0.019}_{-0.017}$  &  $0.774 (0.761) \pm 0.018$  &    $0.8114 (0.8122) \pm 0.0087$     & $0.8035 (0.8070) \pm 0.0081$\\

${\rm{Age}}/\mathrm{Gyr}   $ & $13.823 (13.809)^{+0.027}_{-0.041}$  & $13.951 (14.008)^{+0.062}_{-0.082}$ &  $13.796 (13.773)^{+0.021}_{-0.029}$    & $13.785 (13.808)^{+0.024}_{-0.035}  $\\

$r_\mathrm{drag}$ [Mpc] & $147.96 (147.50)^{+0.31}_{-1.1}$  & $153.3 (155.1)^{+1.8}_{-2.5}$ &  $147.87 (147.55)^{+0.26}_{-0.51}$    & $148.1 (148.3)^{+1.1}_{-0.87}$\\

    \hline

$\chi^2_{\mathrm{bf},Planck}$       & $2764.3$  &  $2772.2$  &   $2764.8$     & $2765.8$\\

$\chi^2_{\mathrm{bf, total}}$       & $2764.3$  &  $2773.2$  &    $3820.9$    & $3838.0$ \\
    
    \hline
  \end{tabular} 
  \caption{Mean (best-fit) $\pm 1\sigma$ constraints on cosmological parameters in the DCDM$\gamma$ model from various dataset combinations.  Sampled parameters are shown in the first nine rows.  Upper and lower limits are given at 95\% CL.  For comparison, a $\Lambda$CDM fit to \emph{Planck} CMB + \emph{FIRAS} yields $r_{\rm drag} = 147.08(147.00) \pm 0.29$ Mpc and $\chi^2_{\mathrm{bf}, Planck} = 2765.5$.}
  \label{table:DCDMgamma}
\end{table*}

\subsection*{DCDM$\gamma$+$N_{\rm eff}$ Analysis}

\begin{figure*}[!tp]
\includegraphics[width=\textwidth]{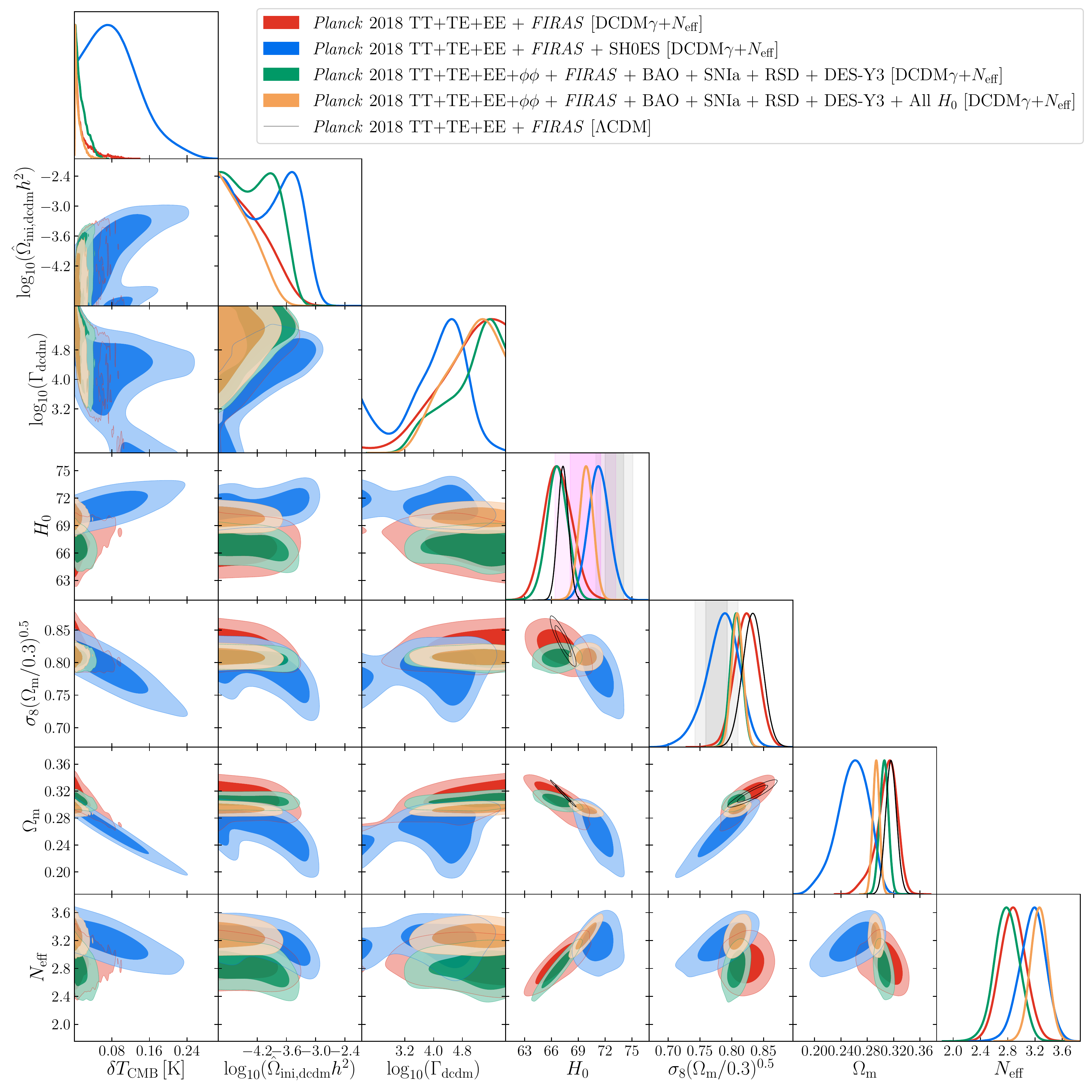}
\caption{Marginalized posteriors for the DCDM$\gamma$+$N_{\rm eff}$ parameters and select other parameters in fits to various datasets, as labeled, with $H_0$ and $\Gamma$ in units of km/s/Mpc..  For comparison, $\Lambda$CDM constraints for \emph{Planck} CMB and \emph{FIRAS} data are shown in thin black lines.  The vertical grey and magenta bands are identical to those in Fig.~\ref{fig:DCDMgamma}.}
\label{fig:DCDMgammaNeff}
\end{figure*}

Here we include results for the DCDM$\gamma$+$N_{\rm eff}$ analysis (Table~\ref{table:DCDMgammaNeff} and Fig.~\ref{fig:DCDMgammaNeff}), considering the same dataset combinations as used to constrain the DCDM$\gamma$ model in the main text.  The inclusion of DR (parameterized by $N_{\rm eff}$) as an additional component in the model is motivated by (1) the fact that the DCDM component itself is likely to be (semi-)relativistic at recombination and (2) the need to mitigate the increase in $\Omega_{\rm m}$ necessary to maintain the value of $k_{\rm eq}$ despite the lower photon energy density in the early universe (see discussion in the main text).  Regarding point (1), strictly speaking the DCDM and DR components may in fact be identified with one another in the context of a specific particle physics model, i.e., it would be more correct to consider a ``DR$\gamma$'' model or a model with decaying warm dark matter, since the idea is that the decaying component is sufficiently light to contribute to $N_{\rm eff}$ at recombination.  In the current implementation, we do not ``remove'' the additional $N_{\rm eff}$ energy density after recombination, i.e., it is simply taken to be a separate new species, in addition to the DCDM component.  An explicit particle physics model may thus have fewer free parameters than considered here.

\begin{table*}[ht!]
Constraints on DCDM$\gamma$+$N_{\rm eff}$ Parameters  \\
  \centering
  %\begin{tabular}{|c|c|c|c|c|c|}
  \begin{tabular}{|c|c|c|c|c|}
    \hline\hline Parameter & \hspace{0cm} 
    \begin{tabular}[t]{@{}c@{}}\emph{Planck} CMB \\ TT+TE+EE, \\ \emph{FIRAS} \end{tabular} & \hspace{0cm}
    \begin{tabular}[t]{@{}c@{}}\emph{Planck} CMB \\ TT+TE+EE, \\ \emph{FIRAS}, SH0ES \end{tabular} & \hspace{0cm}
    %\begin{tabular}[t]{@{}c@{}}\emph{Planck} CMB\\ TT+TE+EE, \\  \emph{FIRAS}, BAO \end{tabular} & \hspace{0cm}
    \begin{tabular}[t]{@{}c@{}}\emph{Planck} CMB\\ TT+TE+EE, \\  \emph{FIRAS}, BAO, \\ \emph{Planck} $\phi\phi$, SNIa, \\ RSD, DES-Y3\end{tabular} &
    \hspace{0cm}
    \begin{tabular}[t]{@{}c@{}}\emph{Planck} CMB\\ TT+TE+EE, \\  \emph{FIRAS}, BAO, \\ \emph{Planck} $\phi\phi$, SNIa, \\ RSD, DES-Y3, \\ Combined $H_0$\end{tabular}
    \\\hline \hline

{$T_\mathrm{CMB,ini}$} [K] & $2.707 (2.723)^{+0.019}_{-0.0020}$  &  $2.638 (2.624)^{+0.068}_{-0.036}$  &  $2.711 (2.713)^{+0.014}_{-0.0045}$   & $2.7162 (2.7182)^{+0.0091}_{-0.0021}$    \\

{$\log_{10}(\hat{\omega}_\mathrm{DCDM,ini})$} &  $< -3.58 (-4.44)$ & $< -3.15 (-3.36)$  &   $< -3.53 (-4.04)$   & $< -3.88 (-4.29)$   \\

{$\log_{10}(\frac{\Gamma}{{\rm km/s/Mpc}})$} & $> 3.30 (5.55)$  & $4.07 (4.76)^{+0.98}_{-0.50}$ & $> 3.59 (5.15)$     & $> 3.74 (5.13)$   \\

{$\ln(10^{10}\hat{A}_\mathrm{s})$} & $3.036 (3.039) \pm 0.018$  &  $3.051 (3.054) \pm 0.019$   & $3.028 (3.036)\pm 0.016$   & $3.051 (3.052)\pm 0.014$    \\

{$n_\mathrm{s}   $} & $0.9583 (0.9572) \pm 0.0085$  & $0.9722 (0.9654)^{+0.0086}_{-0.0074}$  &   $0.9575 (0.9626)^{+0.0081}_{-0.0072}$   & $0.9763 (0.9766) \pm 0.0055$   \\

% formatting compression
%{$100\theta_\mathrm{s}$} & $1.04226 (1.04248) \pm 0.00052$ & $1.04172 (1.04189) \pm 0.00050$   & $1.04255 (1.04235)\pm 0.00052$    & $1.04154 (1.04157)\pm 0.00041$    \\

{$100\theta_\mathrm{s}$} $\times 10^4$ & $10422.6 (10424.8) \pm 5.2$ & $10417.2 (10418.9) \pm 5.0$   & $10425.5 (10423.5)\pm 5.2$    & $10415.4 (10415.7)\pm 4.1$    \\

% formatting compression
%{$\hat{\Omega}_\mathrm{b} h^2$} &$0.02218 (0.02216) \pm 0.00023$  &  $0.02252 (0.02230)^{+0.00024}_{-0.00020}$     &   $0.02219 (0.02234)^{+0.00024}_{-0.00020}$    & $0.02267 (0.02270) \pm 0.00016$   \\

{$\hat{\Omega}_\mathrm{b} h^2$} $\times 100$ &$2.218 (2.216) \pm 0.023$  &  $2.252 (2.230)^{+0.024}_{-0.020}$     &   $2.219 (2.234)^{+0.024}_{-0.020}$    & $2.267 (2.270) \pm 0.016$   \\

{$\hat{\Omega}_\mathrm{c} h^2$} & $0.1176 (0.1161) \pm 0.0031$  & $0.1207 (0.1196) \pm 0.0031$  &   $0.1150 (0.1162)\pm 0.0029$   & $0.1214 (0.1207) \pm 0.0023$   \\

{$\tau_\mathrm{reio}$} & $0.0533 (0.0556) \pm 0.0076$  &  $0.0560 (0.0568)^{+0.0076}_{-0.0085}$  &   $0.0525 (0.0549)\pm 0.0070$    & $0.0556 (0.0579) \pm 0.0072$    \\

{$N_\mathrm{ur}$} & $1.86 (1.78) \pm 0.19$  &  $2.15 (2.02)^{+0.20}_{-0.18}$  &   $1.77 (1.86)\pm 0.18$    & $2.25 (2.22) \pm 0.13$    \\

    \hline

$\delta T_{\rm CMB}$ [K]   &  $< 0.0675 (0.0028)$  & $< 0.193 (0.102)$  &   $< 0.0390 (0.0126)$   & $< 0.0268 (0.0073) $ \\

$H_0       $ [km/s/Mpc]      & $66.7 (65.8) \pm 1.6$  & $71.3 (70.6) \pm 1.2$ &  $66.6 (67.3)\pm 1.2$     & $69.88 (69.78) \pm 0.79$\\

$\Omega_\mathrm{c} h^2     $ &  $0.1152 (0.1158)^{+0.0044}_{-0.0033}$  & $0.1097 (0.1068)^{+0.0097}_{-0.0069}$  &   $0.1131 (0.1146)\pm 0.0034$   & $0.1202 (0.1197) \pm 0.0025$\\

$\Omega_\mathrm{m}         $ & $0.309 (0.320)^{+0.018}_{-0.011}  $   & $0.258 (0.256)^{+0.027}_{-0.020}$ &  $0.3054 (0.3031)\pm 0.0066$     & $0.2934 (0.2934) \pm 0.0051$\\

$\sigma_8                  $ & $0.811 (0.804)^{+0.012}_{-0.016} $  & $0.848 (0.855)^{+0.017}_{-0.022}$   &   $0.8000 (0.8043) \pm 0.0097$     & $0.8178 (0.8157) \pm 0.0082$\\

$S_8$                        & $0.823 (0.829)\pm 0.018$  &  $0.785 (0.789)^{+0.029}_{-0.022}$  &    $0.8071 (0.8084) \pm 0.0094$     & $0.8088 (0.8067) \pm 0.0090$\\

${\rm{Age}}/\mathrm{Gyr}   $ & $14.01 (14.06) \pm 0.21$  & $13.80 (13.96)^{+0.21}_{-0.30}$ &  $14.08 (13.97) \pm 0.20$    & $13.57 (13.59) \pm 0.13 $\\

$r_\mathrm{drag}$ [Mpc] & $149.9 (149.8)^{+2.0}_{-2.5}$  & $151.0 (152.9)^{+3.3}_{-5.4}$ &  $150.8 (149.7)^{+2.0}_{-2.4}$    & $145.8 (146.0)^{+1.3}_{-1.5}$\\

$N_{\rm eff}$                        & $2.87 (2.79) \pm 0.19$  &  $3.17 (3.04)^{+0.20}_{-0.18}$  &    $2.79 (2.88) \pm 0.18$     & $3.26 (3.23) \pm 0.13$\\

    \hline

$\chi^2_{\mathrm{bf},Planck}$     & $2763.3$   &  $2769.4$  &    $2764.3$     & $2769.0$\\

$\chi^2_{\mathrm{bf, total}}$     & $2763.3$   &  $2773.5$  &    $3819.4$     & $3834.3$ \\
    
    \hline
  \end{tabular} 
  \caption{Mean (best-fit) $\pm 1\sigma$ constraints on cosmological parameters in the DCDM$\gamma$+$N_{\rm eff}$ model from various dataset combinations.  Sampled parameters are shown in the first ten rows.  Upper and lower limits are given at 95\% CL.  The $\chi^2_{\mathrm{bf, total}}$ for \emph{Planck} CMB + \emph{FIRAS} + SH0ES is slightly higher than that found in Table~\ref{table:DCDMgamma} because of small improvements in the fits to the \emph{Planck} nuisance parameter priors, whose $\chi^2$ contributions are not included here.  For comparison, a $\Lambda$CDM fit to \emph{Planck} CMB + \emph{FIRAS} yields $r_{\rm drag} = 147.08(147.00) \pm 0.29$ Mpc and $\chi^2_{\mathrm{bf}, Planck} = 2765.5$, while a $\Lambda$CDM+$N_{\rm eff}$ fit to the full combination of datasets considered in the rightmost column yields $H_0 = 70.04(70.28) \pm 0.78$ km/s/Mpc, $S_8 = 0.8101(0.8098) \pm 0.0090$, $r_{\rm drag} = 144.9(144.6) \pm 1.2$ Mpc, $\chi^2_{\mathrm{bf}, Planck} = 2771.5$, and $\chi^2_{\mathrm{bf, total}} = 3834.8$.}
  \label{table:DCDMgammaNeff}
\end{table*}

As for DCDM$\gamma$, primary CMB data alone do not show a strong preference for the DCDM$\gamma$+$N_{\rm eff}$ model over $\Lambda$CDM: we obtain $\Delta \chi^2 = -2.2$ in favor of DCDM$\gamma$+$N_{\rm eff}$.  The inclusion of SH0ES $H_0$ data yields a mild preference for non-zero DCDM, but less strong than in the (zero-$N_{\rm eff}$) DCDM$\gamma$ model, as here $N_{\rm eff}$ absorbs some of the change (although note that $r_s^\star$ is still increased here, relative to its $\Lambda$CDM value).  Formally, the \emph{Planck} CMB + \emph{FIRAS} + SH0ES analysis yields an upper bound $\log_{10}(\hat{\omega}_\mathrm{DCDM,ini}) < -3.15$ at 95\% CL, with $H_0 = 71.3 \pm 1.2$ km/s/Mpc, $S_8 = 0.785^{+0.029}_{-0.022}$, and $\delta T_{\rm CMB} < 0.193$~K, with a long tail extending to larger values than allowed in the analogous DCDM$\gamma$ analysis (as anticipated by the inclusion of $N_{\rm eff}$).  The fit to \emph{Planck} CMB data is only mildly degraded, with $\Delta \chi^2_{Planck} = 3.9$ compared to the best-fit $\Lambda$CDM model to \emph{Planck} alone.

Removing SH0ES but including the full complement of non-$H_0$ datasets, the hint for DCDM weakens but is notably stronger than for primary CMB data alone in the DCDM$\gamma$+$N_{\rm eff}$ analysis; here the DES-Y3 $S_8$ constraint plays an important role.  However, the central value of $H_0$ does not increase compared to its primary-CMB value, despite the reduction in $S_8$.

With all datasets combined (including all three independent $H_0$ probes), we find $H_0 = 69.88 \pm 0.79$ km/s/Mpc, $S_8 = 0.8088 \pm 0.0090$, and $\delta T_{\rm CMB} < 0.0268$~K (95\% CL upper limit).  The Hubble constant is thus increased substantially more than in DCDM$\gamma$ (where the analogous analysis yields $H_0 = 68.69 \pm 0.35$ km/s/Mpc), due to the role played by $N_{\rm eff}$ here.  The error bar is also significantly increased, due to the parameter degeneracies in the radiation sector of this model.  However, the fit to \emph{Planck} is slightly degraded compared to that of $\Lambda$CDM, with $\Delta \chi_{Planck}^2 = 3.5$; this very likely arises due to the \emph{Planck} polarization data disfavoring large values of $N_{\rm eff}$~\cite{Planck2018parameters}.  Note that the role played by the DCDM component here in increasing $H_0$ is now sub-dominant to the role played by $N_{\rm eff}$ via the standard sound-horizon reduction mechanism~\cite{Knox:2019rjx}, as can be seen in the preferred values of $r_{\rm drag}$, which now lie below the best-fit $\Lambda$CDM value.

For a final comparison, we fit a $\Lambda$CDM+$N_{\rm eff}$ model to the full suite of datasets.  Interestingly, we obtain $H_0$ values very similar to those found in the DCDM$\gamma$+$N_{\rm eff}$ fit, with $H_0 = 70.04 \pm 0.78$ km/s/Mpc, but the fit to the \emph{Planck} CMB data is somewhat more degraded, with $\Delta \chi_{Planck}^2 = 6.0$.  The relative improvement of the \emph{Planck} CMB fit in DCDM$\gamma$+$N_{\rm eff}$ compared to $\Lambda$CDM+$N_{\rm eff}$ is thus $\Delta \chi_{Planck}^2 = -2.5$ in favor of the former.  While this is not significant, it does indicate that the DCDM component helps in allowing the $N_{\rm eff}$ sound-horizon reduction mechanism to operate without degrading the fit to \emph{Planck} as much as it otherwise would.

As in the DCDM$\gamma$ analysis, although the DCDM component is not detected, all dataset combinations disfavor low values of $\Gamma$ (i.e., low $z_X$).  In the \emph{Planck}+\emph{FIRAS}+SH0ES fit to DCDM$\gamma$+$N_{\rm eff}$, the best-fit $z_X \simeq 75$; in the joint analysis of all datasets, the best-fit $z_X \simeq 130$ ($z_X \gtrsim 15$ at 95\% CL).  Thus, as found earlier, the (mildly) preferred decay redshifts are prior to reionization, where no direct observational constraints exist.

Our results indicate that small amounts of PRR are allowed, but tightly constrained, by background and linear cosmological observables.  When additional DR is included in the model, the bounds on PRR generally weaken slightly (although actually tighten slightly in the joint fit of all datasets), but still restrict the change in the CMB monopole temperature to $\lesssim 30$-40 mK and $f_{\rm DCDM} \lesssim 0.2$-0.3\%.  These bounds are completely independent from the spectral distortion signatures of such a scenario.

% Figure matching Table 4

%%%%%%%%%%%%%%%%%%%%%%%%%%%%%%%%%%%%%%%%%%%%%%

\bibliographystyle{JHEP}
\bibliography{refs}
%TC:endignore

\end{document}